\documentclass[11pt,a4paper]{article}

\usepackage[T1]{fontenc}
\usepackage{graphicx}
\usepackage{microtype}
\usepackage{epsfig}
\usepackage{subcaption}
\usepackage{calc}
\usepackage{amssymb}
\usepackage{amstext}
\usepackage{amsmath}
\usepackage{amsthm}
\usepackage{multicol}
\usepackage{pslatex}
\usepackage{algorithm2e}

\usepackage{amsmath,amssymb,amsfonts}
\usepackage{algorithmic}
\usepackage{graphicx}
\usepackage{textcomp}
\usepackage{comment}
\usepackage{xcolor}

\usepackage{tikz}
\usetikzlibrary{positioning,automata,arrows,calc,shapes,graphs,graphs.standard,quotes}
\tikzset{
modal/.style={>=stealth’,shorten >=1pt,shorten <=1pt,auto,node distance=1.5cm,
semithick},
world/.style={circle,draw,minimum size=0.5cm,fill=gray!15},
point/.style={circle,draw,inner sep=0.5mm,fill=black},
reflexive above/.style={->,loop,looseness=7,in=120,out=60},
reflexive below/.style={->,loop,looseness=7,in=240,out=300},
reflexive left/.style={->,loop,looseness=7,in=150,out=210},
reflexive right/.style={->,loop,looseness=7,in=30,out=330}
}

\newcommand{\healthyc}{\mathfrak{h}}
\newcommand{\mediumc}{\mathfrak{m}}
\newcommand{\strongc}{\mathfrak{M}}
\newcommand{\prob}{P}

\newcommand{\A}{\mathcal{A}}

\newcommand{\hhh}{{\mathfrak{h}}}
\newcommand{\mmm}{{\mathfrak{m}}}
\newcommand{\sss}{{\mathfrak{M}}}
\newcommand{\Ah}{_{\mathfrak{h}}}
\newcommand{\Am}{_{\mathfrak{m}}}
\newcommand{\As}{_{\mathfrak{M}}}

\newcommand{\C}{\mathcal{C}}

\newcommand{\M}{\mathcal{M}}

\newcommand{\bel}{\mathcal{B}}
\newcommand{\V}{\mathcal{V}}

\newcommand{\metq}{\mathcal{Q}}

\newcommand{\metlang}{\mathcal{L}_{\mathcal{A}}}

\usepackage{pst-all}
\usepackage{pstricks-add}

\theoremstyle{definition}
\newtheorem{definition}{Definition}

\begin{document}
\title{A Probabilistic Model-Checking Framework for\\ Cognitive Assessment and Training}

\author{
Elisabetta De Maria\\
\small Université Côte d'Azur, I3S, CNRS, France\\
\small \texttt{elisabetta.de-maria@univ-cotedazur.fr}
\and
Christopher Leturc\\
\small Université Côte d'Azur, I3S, CNRS, France\\
\small \texttt{christopher.leturc@univ-cotedazur.fr}
}

\date{}
\maketitle             
\begin{abstract}
Serious games have proven to be effective tools for screening cognitive impairments and supporting diagnosis in patients with neurodegenerative diseases like Alzheimer’s and Parkinson’s. They also offer cognitive training benefits. According to the DSM-5 classification, cognitive disorders are categorized as Mild Neurocognitive Disorders (mild NCDs) and Major Neurocognitive Disorders (Major NCDs). In this study, we focus on three patient groups: healthy, mild NCD, and Major NCD. We employ Discrete Time Markov Chains to model the behavior exhibited by each group while interacting with serious games. By applying model-checking techniques, we can identify discrepancies between expected and actual gameplay behavior. The primary contribution of this work is a novel theoretical framework designed to assess how a practitioner’s confidence level in diagnosing a patient’s Alzheimer’s stage evolves with each game session (\emph{diagnosis support}). Additionally, we propose an experimental protocol where the difficulty of subsequent game sessions is dynamically adjusted based on the patient’s observed behavior in previous sessions (\emph{training support}).

\end{abstract}
\section {Introduction}

Neurodegenerative diseases such as Alzheimer’s and Parkinson’s are often associated with the progressive decline of cognitive functions, leading to significant cognitive impairments. Detecting neurocognitive disorders early and continuously monitoring their progression are critical for timely and individualized intervention. Currently, accurate diagnosis typically requires a comprehensive series of neuropsychological assessments, often supplemented by biomarker tests. However, these procedures can be resource-intensive and time-consuming for both clinicians and patients.

Given these challenges, there is growing interest in identifying objective and easily administered behavioral markers that can complement conventional clinical assessments, facilitating early detection of cognitive decline. In this context, \emph{serious games} have emerged as a promising tool \cite{R2014FAN}. These are digital or physical games designed with a purpose beyond entertainment, such as education, training, or health promotion, while retaining engaging gameplay elements~\cite{A007ALE}.

Several feasibility studies have underscored the potential of serious games in assessing cognitive impairments, particularly in the context of neurodegenerative diseases \cite{TCTL16JMIR,V2017PLOS,KK17JATT}. Moreover, they have demonstrated significant value in Alzheimer’s disease therapy by serving as cognitive training tools that can help maintain or even improve cognitive functions \cite{ABR13Nat2013,J2014AJAD}.

Additionally, studies suggest that older adults tend to prefer game-based interventions over traditional cognitive exercises, potentially enhancing both engagement and adherence to therapeutic protocols~\cite{M2006PA}. This preference, combined with the potential diagnostic and training capabilities of serious games, positions them as a compelling, dual-purpose tool for both cognitive assessment and rehabilitation in neurodegenerative conditions.

According to the latest DSM-5 classification \cite{APA2013APA}, cognitive impairments encompass both cognitive decline and behavioral disturbances that can significantly interfere with daily life. Based on the severity of these deficits and their impact on daily functioning, the DSM-5 differentiates between Mild Neurocognitive Disorder (mild NCD) and Major Neurocognitive Disorder (Major NCD). Patients diagnosed with either mild or major NCD typically require ongoing monitoring and support from medical practitioners to manage symptoms and maintain quality of life.

In this paper, which extends the work introduced in \cite{healthinf25}, we advocate for the use of serious games as a complementary tool for both screening cognitive deficits and providing targeted training activities tailored to each patient’s cognitive state. For specific games, multiple difficulty levels can be configured to suit varying levels of cognitive impairment.

Our methodology relies on Discrete Time Markov Chains (DTMCs) to model patient behavior during gameplay, drawing on approaches similar to those presented in \cite{DLMR19FTSCS,LD2021SCP}. When engaging with serious games, patients tend to exhibit recurring patterns of behavior, with some actions being more frequent while others occur less often. We quantify these behavioral variations by associating probabilities with key in-game actions. These probabilities vary based on the patient’s Alzheimer’s severity level (healthy, mild NCD, Major NCD).

Initially, the model assigns \emph{a priori} probabilities to each action based on practitioner input. These probabilities are derived from clinical experience and are subsequently refined using empirical data collected during gameplay sessions. This iterative process ensures that the model accurately reflects patient behavior and can adapt to emerging clinical findings.

For the successful implementation of our framework, practitioners must provide two essential inputs:
\begin{enumerate}
\item    For each game and each Alzheimer’s degree, the \emph{a priori} probabilities associated with significant in-game actions. Whenever available, data-driven probabilities derived from clinical studies will be used to replace or adjust these initial weights.
\item    The initial hypothesis regarding each patient’s cognitive impairment level, based on a preliminary assessment using conventional neuropsychological tests.
\end{enumerate}
To ensure the robustness and validity of our methodology, we employ probabilistic model checking \cite{Hansson94}. This formal verification technique enables the automatic evaluation of model properties, such as confirming that all game executions eventually reach a designated final state. Additionally, model checking facilitates the comparison between expected and observed behavior by computing probabilities for various execution traces. This step is crucial for identifying discrepancies and refining diagnostic or training protocols based on the patient’s performance trajectory.

A key contribution of this work is the introduction of a meta-automaton framework, where each node represents a Markov chain that models the expected behavior of a specific patient group (e.g., healthy, mild NCD, Major NCD) for a given game. Each state within the meta-automaton corresponds to a distinct game session, enabling the analysis of patient behavior over multiple sessions. This structure serves two primary purposes:

\begin{enumerate}
    \item The meta-automaton facilitates the assessment of how a practitioner’s confidence level in a patient’s Alzheimer’s severity evolves after each game session. By analyzing the transitions between states, the framework can indicate whether the patient’s observed behavior aligns with the initial diagnostic classification or suggests a potential reclassification. For instance, a patient initially labeled as “Major NCD” could, based on improved gameplay performance, be considered for reclassification as “mild NCD,” and vice versa.
    \item The framework also supports the implementation of a dynamic experimental protocol in which the difficulty level of subsequent game sessions is automatically adjusted based on observed behavior in previous sessions. This adaptive mechanism accounts for improvements, regressions, or consistent patterns in gameplay, ensuring that each session is tailored to the patient’s evolving cognitive profile.
\end{enumerate}

The protocol concludes once specific terminal conditions are met, such as reaching a predefined number of sessions or achieving a stable confidence level regarding the patient’s cognitive status. Additionally, model checking techniques can detect potential instabilities in patient performance, such as oscillating patterns in behavior that may warrant closer examination.

Model checking~\cite{CEGOPD99MIT} is thus employed at two distinct levels:
to validate the Markov chains that represent patient behavior during serious games, ensuring that the expected sequences of actions are logically consistent and complete, and to assess the properties of the execution traces of the meta-automaton, particularly those related to diagnostic confidence and session outcomes.
To effectively represent the evolving confidence or belief states of practitioners regarding the patient’s Alzheimer’s severity, we advocate the use of doxastic logic. This formalism allows us to express and analyze nuanced changes in belief states as new information is gathered through successive game sessions.

The structure of the paper is as follows. Section \ref{sec:case-study} presents a simple serious game developed at Claude Pompidou Institute (Nice, France), serving as a case study to illustrate the proposed approach. Section \ref{sec:prel} provides the necessary preliminaries on the formal tools employed, including probabilistic automata and temporal logics. The formal framework designed to assist in the diagnosis and cognitive training of Alzheimer’s patients is detailed in Section \ref{sec:framework}. A formal validation of the proposed methodology is presented in Section \ref{sec:disc}, followed by a discussion of key future developments in Section \ref{sec:conclu}.

\section{Case study}~\label{sec:case-study}
\begin{figure}
    \centering
    \includegraphics[scale=0.22]{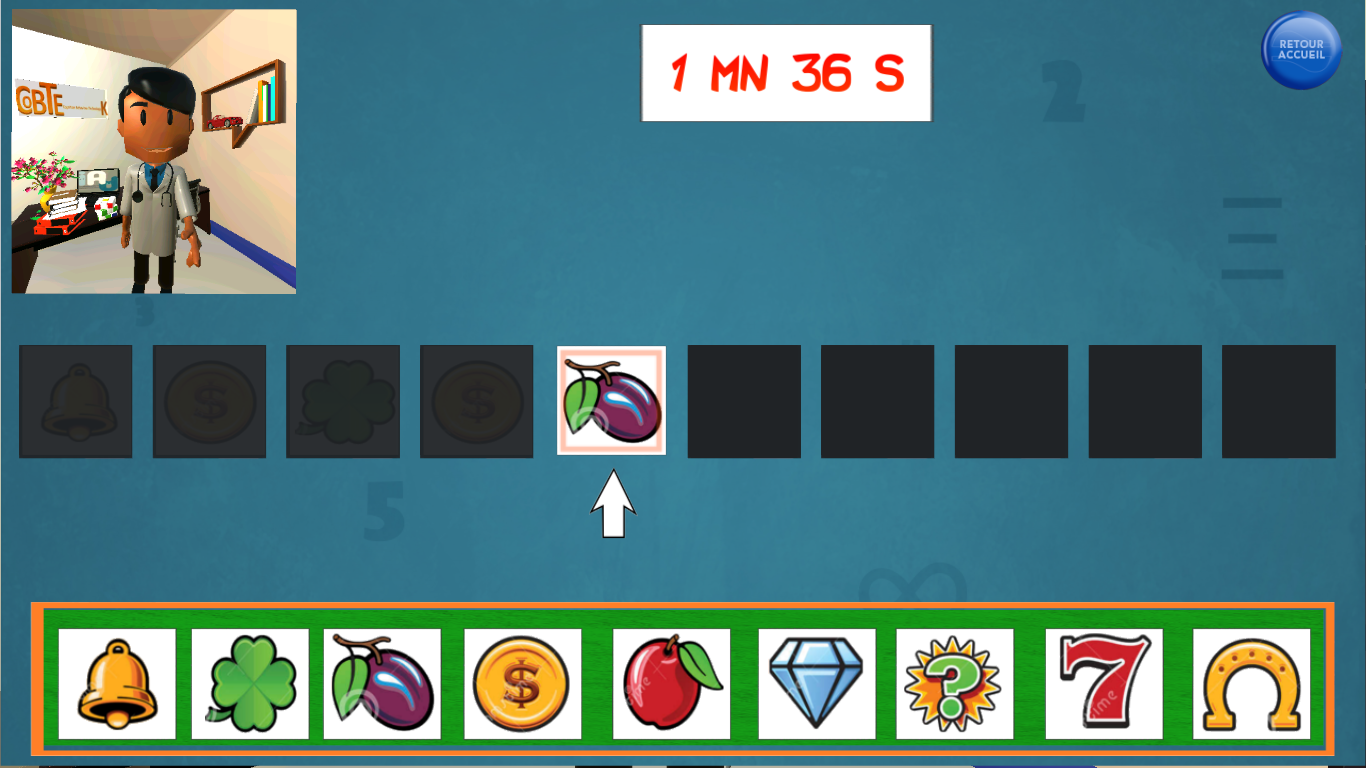}
    \caption{Screenshot of the Match Items game \cite{healthinf25}} 
    \label{fig:match}
\end{figure}

As a simple case study, we introduce the Match Items game~\cite{PBR15gala}, which has been developed in 2014 at Claude Pompidou Institute (Nice, FRance)  and has already been implied in several clinical protocols. In particular, one of the clinical experiments validated this game as a suitable tool to discriminate between mild NCD patients and healthy ones. The Match Items game targets selective
and sustained visual attention functions. In this game  patients interact with a touch-pad. Their task consists in matching a random picture shown at the center of the touch-pad with its corresponding element from a list of pictures located at the bottom of the screen (see Figure~\ref{fig:match}). The game
lasts at most five minutes. 

When the patient selects the correct picture, a happy smiley appears, and a new picture is displayed. In case of an incorrect choice, a sad smiley is displayed, prompting the patient to try again. If there is over 10 seconds of inactivity on the touch-pad, the game reminds the patient to select a picture. Exiting the game zone results in the game being stopped. In the rest of the paper, key actions of this game will be referred as follows: $\alpha :=$ "the patient chooses the right picture", $\beta :=$ "the patient chooses the wrong picture", $\gamma :=$ "the patient is inactive", $\theta :=$ "the patient quits the game zone".

\section{Preliminaries}~\label{sec:prel}

In this section, we introduce Probabilistic Finite Deterministic Automata (PDFAs) and temporal logics. These models are used to formalise and analyse the probabilistic behaviour of players in a game. Specifically, PDFAs provide a framework for representing players' actions in the game, while temporal logics allow us to express the temporal properties and constraints of their behaviour.

\subsection{Probabilistic Deterministic Finite Automata}

Let us consider patients engaged in playing serious games. 
Given our focus on studying the patients activity and the unpredictability of patients' actions, we adopt Probabilistic Deterministic Finite Automata (PDFA)~\cite{rabin1963probabilistic}. Given a serious game, we conceive an automaton for each class of patients. This automaton serves a dual purpose by  representing the development of the serious game and specifying the activity the user is supposed to display if she belongs to the class being tested by the automaton.

Each serious game is represented as a PDFA, where the states represent the different game configurations, e.g., the user has to choose a picture, or the end of the game. The input symbols of the alphabet represent the actions the user can make, denoted by $\Sigma_g = \{\alpha, \beta, \gamma, \theta\}$. We note $\Sigma_g^* \subseteq \Sigma_g^{\mathbb{N}^*}$ the set of words  we can write with $\Sigma_g$. A word is just a concatenation of symbols on  $\Sigma_g$, $w = \alpha \beta \gamma \beta \gamma \beta\alpha\beta \theta$ is a word that uses the symbols of $\Sigma_g$, thus $w \in \Sigma_g^*$.

\begin{definition}~\label{def:dfa}
    A \emph{Deterministic Finite Automaton} (DFA) is a 5-tuple $A = (Q, \Sigma, \delta_{d}, q_0, F)$ where:
\begin{itemize}
    \item $Q$ is a finite set of states.
    \item $\Sigma$ is a finite alphabet of input symbols.
    \item $\delta_{d}: Q \times \Sigma  \rightarrow Q$ is the  transition function, where $\delta_{d}(q, a)$ represents the next state when being in state $q$ and reading input symbol $a$.
    \item $q_0 \in Q$ is the initial state.
    \item $F \subseteq Q$ is the set of accepting (final) states.
\end{itemize}

We define $\delta^*$ the transition function w.r.t. $\delta$ which operates on a set of words:
\begin{align}
    \delta^*_{d}: & Q \times \Sigma^* \rightarrow Q \\
    & \delta^*_{d}(q, ()) = q \\
    & \delta^*_{d}(q, xa) = \delta_{d}(\delta^*_{d}(q, x), a) \quad \text{for } x \in \Sigma^*, \text{ } a \in \Sigma.
\end{align}

A language $L_{d}$ recognized by a DFA is defined as:
\[
L_{d} =  \{w \in \Sigma^* \mid \delta^*_{d}(q_0, w) \in F\}
\]

We say that $A = (Q, \Sigma, \delta_{d}, \prob, q_0, F)$ is a \emph{Probabilistic Deterministic Finite Automaton} (PDFA)  if and only if, $A = (Q, \Sigma, \delta_{d}, q_0, F)$ is a DFA and $\prob: Q \times \Sigma \times Q \rightarrow [0,1]$ is a probabilistic function such that : 
\begin{align*}
    (1)\; & \forall (q,a,q') \in  Q \times \Sigma \times Q, \text{ if } \delta_{d}(q,a) \neq q'\\ & \text{ then }  \prob(q,a,q') = 0 \\
    (2)\; & \forall q \in  Q , \sum\limits_{(a,q') \in \Sigma \times Q} \prob(q,a,q') = 1 \\
\end{align*}

\end{definition}

When the future states of a PDFA depend only on the present state and are independent of the sequence of events that preceded it,  the Markov property holds~\cite{norris1998markov}. In other words, given the present, the past has no additional information to offer about the future. A PDFA with the Markov property is called a \emph{Markov chain}. 

 The use of Markov chains provides a tool for modeling the probabilistic behavior of patients while playing serious games. In the sequel, we present temporal logics, a formal technique which is used to automatically analyze the probabilistic behavior of patients and verify the validity of specific properties while they are playing.

\subsection{Temporal Logics}
Temporal logic formulae describe the dynamical evolution of a given system. The \emph{Computation Tree Logic}
CTL$^\ast$ \cite{CEGOPD99MIT} allows one to describe properties of computation trees. Its formulas are obtained by (repeatedly) applying Boolean connectives, \emph{path quantifiers}, and \emph{state quantifiers} to atomic formulas. The path quantifier $\textbf{A}$ (resp., $\textbf{E}$) can be used to state that all paths (resp., some path) starting from a given state have some property. The state quantifiers are the following ones. The next time operator $\textbf{X}$ can be used to impose that a property holds at the next state of a path. The operator $\textbf{F}$ (sometimes in the future) requires that a property holds at some state on the path. The operator $\textbf{G}$ (always in the future) specifies that a property is true at every state on the path. The until binary operator $\textbf{U}$ holds if there is a state on the path where the second of its argument properties holds, and, at every preceding state on the path, the first of its two argument properties holds. The \emph{Branching Time Logic} CTL \cite{CES86TPLS} is a fragment of CTL$^\ast$ that allows quantification over the paths starting from a given
state. Unlike CTL$^\ast$, it constrains every state quantifier to be immediately preceded by a path quantifier. 
The \emph{Linear Time Logic} LTL \cite{SC85ACM} is another known fragment of CTL$^\ast$ where one may only
describe events along a single computation path. 
Its formulas are of the form $\textbf{A} \varphi$, where $\varphi$ does not contain path quantifiers, but it allows the nesting of state quantifiers. 
CTL and LTL have a non-empty intersection. As an example, the property $\texttt{A ((x=1) U (y=3))}$ belongs both to CTL and LTL. It holds in a state if, for each path starting from the state, $x$ equals $1$ until the moment when $y$ equals 3. 
There exists several tools to automatically check whether a  model verifies a given CTL or LTL formula, e.g., NuSMV \cite{CCGR99CAV} and SPIN \cite{H04Addison}.

The dynamics of probabilistic models can be specified using Probabilistic Computation Tree Logic (PCTL) \cite{Hansson94}, which extends CTL 
    by replacing the classical CTL path quantifiers
    $\textbf{A}$ and $\textbf{E}$ with
    probabilities. Thus, instead of saying that some property holds for all
    paths or for some paths, we say that a property holds for a certain fraction
    of the paths.
    The most important operator in PCTL is $\textbf{P}$, which allows to reason
    about the probability of event occurrences. The property $\textbf{P}\
    \textit{bound}\ \texttt{[}\textit{prop}\texttt{]}$ is true in a state $s$ of
    a model if the probability that the property $prop$ is satisfied by the
    paths from state $s$ satisfies the bound $bound$. As an example, the PCTL
    property $\texttt{P =0.4 [X (y = 2)]}$ holds in a state if the probability
    that $y=2$ is true in the next state equals $0.4$. 
    To compute the likelihood that some behavior of a model happens, the
    $\textbf{P}$ operator can take the form $\texttt{P=?}$. As an example, the
    property $\texttt{P =? [G (y = 1)]}$ assesses the probability that
    $y$ always equals $1$. 
Several model-checkers allow to automatically check whether a given probabilistic model satisfies a given PCTL formula, or to automatically compute the probability for a given formula to be satisfied. State-of-the-art probabilistic model checkers are PRISM~\cite{KNP11CAV}, UPPAAL~\cite{tutorial04},
STORM~\cite{dehnert2017storm}, and PAT~\cite{SunLDP09}.

\section{The framework: a doxastic meta-automaton} ~\label{sec:framework}

In this section, we present the framework while instantiating it to our medical application. First, we define three PDFA to model the behaviour of three classes of patients while playing the game.  
Secondly, we define a 'meta-automaton', which aims to model the protocol. This meta-automaton suggests to practitioners which class the patient is supposed to belong to and helps the patient with training. 
After a game session, based on patient performances, the meta-automaton suggests a class the patient is supposed to belong and the next game session for the patient. 
For the sake of compactness, in the following, we denote the healthy class with $\healthyc$, the mild NCD with $\mediumc$, and the Major NCD with $\strongc$. In the following, each PDFA is considered as a \emph{test} the patient is submitted to. 

\subsection{Three PDFA for three classes of patients}

In the case study concerning the Match Items game (see Section~\ref{sec:case-study}), we consider the following finite alphabet $\Sigma_g = \{\alpha,\beta,\gamma,\theta\}$, which represents the different possible actions defined in Section~\ref{sec:case-study}. We recall that such alphabet represents the following actions : 
\begin{itemize}
    \item $\alpha$ := "the patient chooses the right picture"
    \item $\beta$ := "the patient chooses the wrong picture"
    \item $\gamma$ := "the patient is inactive"
    \item $\theta$ := "the patient quits the game zone"
\end{itemize}

As an example, a word as $w_1 = \alpha\beta\beta\alpha \in \Sigma^{*}_g$ signifies that the user first does action $\alpha$, then $\beta$, then $\beta$, and finally $\alpha$. Let us notice that this notation is equivalent with $w_1 = (\alpha,\beta,\beta,\alpha)$ which is a 4-tuple, or $w_1 = (u_n)_{n \in \{1,2,3,4\}}$ a sequence such that $u_1 = \alpha$, $u_2 = \beta$, $u_3 = \beta$ and $u_4 = \alpha$.

In our medical application, we consider three deteministic automata $\metq = \{A\Ah,A\Am,A\As\}$ since there are three classes of Alzheimer patients. Each automaton represents the activity of a class of patients while playing the serious game. 
In order to validate or reject one hypothesis about a state of a patient, we consider that  $A\Ah$ represents the test for $\healthyc$, $A\Am$ for $\mediumc$, and $A\As$ for $\strongc$. In these automata we consider one initial state $q_0$ in which the user has to launch the game, and two final states: $f_1$ when the game is over, and $f_2$ when the user left the game before it was over. Let $F = \{f_1,f_2\}$ be the final states. We define the following three  PDFAs: for all $x \in \{\sss, \mmm, \hhh\}$, $A_x = (Q_x \cup F, \Sigma_g, \delta_x, P_x, q_x, F)$, where $L_x$ is the language recognized by $A_x$.

For the Match Items game, clinicians already provided us with (a priori) empirical  probabilities on the different actions to be performed depending on the different classes. To obtain these probabilities, 10 clinicians—including medical doctors, nurses, and psychologists who are familiar with patients' performance while playing the game—each filled out a questionnaire. The questionnaire included questions such as: "For a patient in a given class, what are the chances of selecting the correct image at each step?" and "For a patient in a given class, what are the chances of not interacting with the game for at least 10 seconds?" The responses were given as numbers from $0$ to $10$. Table~\ref{tab:avg-prob} represents the average probability given by 10 clinicians.  We assume that, for each of the three automata, the probabilistic function follows this table, e.g., $P\Ah$ is such that  for all $(q,q') \in (Q\Ah\setminus F)^2, P\Ah(q,\alpha,q') = 0.8, P\Ah(q,\beta,q')=0.1, P\Ah(q,\gamma,q')=0.05, P\Ah(q,\theta,q')=0.05$.

\begin{table}
    \centering
    \begin{tabular}{|c|c|c|c|}
     \hline
    Action & $\healthyc$ & $\mediumc$ & $\strongc$ \\
     \hline
     $\alpha$ & 0.84 & 0.5 & 0.17 \\
     \hline
      $\beta$ & 0.11 & 0.30 & 0.58 \\
     \hline
    $\gamma$& 0.0499 & 0.1999 & 0.24\\
     \hline
     $\theta$ & 0.0001  & 0.0001  & 0.01 \\
     \hline
    \end{tabular}
    \caption{Average probability given by 10 clinicians for each class of patients \cite{healthinf25}}
    \label{tab:avg-prob}
\end{table}

\begin{figure}\label{fig:dfa-example}
    \centering
    \begin{tikzpicture}[auto, node distance=2cm, on grid, >=stealth, state/.style={circle, draw, minimum size=1cm}]
    \node[state, initial, initial text={~}] (q0) {$q_{0}$};
    \node[state, right=of q0, initial where=left] (q2) {$q_{11}$};
    \node[state, above=of q2] (q1) {$q_{10}$};

    \node[state, below=of q2] (q3) {$q_{12}$};

    \node[state, right=of q3] (q31) {$q_{121}$};
    \node[state, right=of q2] (q21) {$q_{111}$};
    \node[state, right=of q1] (q11) {$q_{101}$};

    \node[state, right=of q21,accepting] (qf1) {$f_1$};

    \node[state, below=of q3,accepting] (q4) {$f_2$};

    \path[->] (q0) edge[bend left,green] node {$\alpha$} (q1);
    \path[->] (q1) edge[bend left,green] node {$\alpha$} (q11);
    \path[->] (q1) edge[bend left,red] node {$\beta$} (q21);

    \path[->] (q2) edge[bend left,green] node {$\alpha$} (q11);
    \path[->] (q2) edge[bend left,red] node {$\beta$} (q21);

    \path[->] (q3) edge[bend left,green] node {$\alpha$} (q1);
    \path[->] (q3) edge[bend left,red] node {$\beta$} (q2);

    \path[->] (q31) edge[bend left,green] node {$\alpha$} (q1);
    \path[->] (q31) edge[bend left,red] node {$\beta$} (q2);

    \path[->] (q11) edge[bend left] node {$\ldots$} (qf1);
    \path[->] (q21) edge[bend left] node {$\ldots$} (qf1);

    \path[->] (q0) edge[bend left,red] node {$\beta$} (q2);

    \path[->] (q0) edge[bend right] node {$\gamma$} (q3);

    \path[->] (qf1) edge [loop below] node {$\forall$} (qf1);
    \path[->] (q4) edge [loop left] node {$\forall$} (q4);



    \path[->] (q0) edge[bend right] node {$\theta$} (q4);
    \path[->] (q1) edge[bend right] node {$\theta$} (q4);
    \path[->] (q3) edge[bend right] node {$\gamma$} (q31);
    \path[->] (q2) edge[bend right] node {$\theta$} (q4);
    \path[->] (q3) edge[bend right] node {$\theta$} (q4);
    \path[->] (q31) edge[bend right] node {$\theta$} (q4);

    \path[->] (q11) edge[bend left=90] node {$\theta$} (q4);
    \path[->] (q21) edge[bend left=60] node {$\theta$} (q4);

    \end{tikzpicture}
    \caption{Automaton $A\Ah$ describing the expected behaviour of healthy people while playing the Match Items serious game \cite{healthinf25}}
    \label{fig:enter-label}
\end{figure}
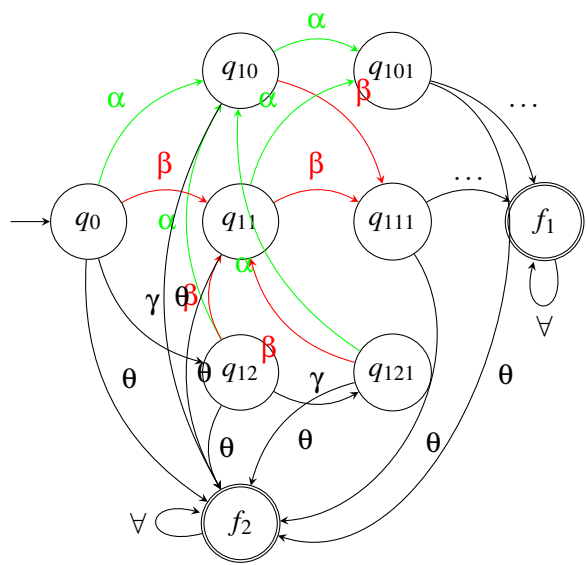
For the sake of clarity, Figure~\ref{fig:enter-label} illustrates the automaton $A\Ah$ in a simplified manner by depicting each transition between states for each action. The initial state is denoted as $q_0$, and there are two final states: $f_1$ signifies the normal end of the game, while $f_2$ indicates the user left the game. Furthermore, let us notice that $A\Ah$ has the Markov chain property as $A\Am$ and $A\As$.

\subsection{An experimental protocol as a meta-automaton}
\label{subsec:ddfma}

The experimental protocol aims to monitor and assess the patient's condition. This protocol involves organizing various tests within a meta-automaton.  
After each test, a \emph{belief function} provides a confident score about the class the patient belongs to. Thanks to this score, the meta-automaton informs the decision-making process for the next test to apply.

\begin{definition}~\label{def:dfa}
    A \emph{Doxastic Deterministic Finite Meta-Automaton} (DDFMA) is a 7-tuple $\A = (\metq,\Sigma_Q, \Sigma_A, \delta, \{\bel_q\}_{q \in Q}, q_0,F)$ where\footnote{To distinguish between PDFA and DDFMA, we denote the latter with a calligraphic letter $\A$}:
\begin{itemize}
    \item $\metq$ is a finite set of PDFA.
    \item $\Sigma_Q = \bigcup\limits_{q \in Q} \Sigma_q$ is a finite alphabet of input symbols recognized by all automata in $Q$.
    \item $\Sigma_A$ is a finite alphabet of input symbols.
    \item $\delta: \metq \times \Sigma_A  \rightarrow \metq$ is the transition function. 
    \item $\forall q \in \metq,  \bel_q:L_q \times \Sigma_A \times \metq \rightarrow [0,1]$ is a belief function that represents from an automaton $q$, given an accepted word $w \in L_q$, the practitioner belief for the patient to be in the class associated with the next automaton $q'$. Furthermore $\bel_q$ is such that: 
    \begin{align*}
        (1)\; & \forall q \in \metq, \forall (a,q') \in  \Sigma_A \times \metq, \text{ if } \delta (q,a) \neq q' \\ 
        & \text{ then } \forall w \in L_q, \bel_q(w,a,q') = 0 \\
        (2)\; & \forall q \in \metq, \forall w \in L_q, \sum\limits_{(a,q') \in \Sigma_A \times \metq} \bel_q(w,a,q') = 1     
    \end{align*}
    \item $q_0 \in \metq$ is the initial state, i.e., the first automaton.
    \item $F$ is the set of final states.
\end{itemize}

We define $\delta^*$ the transition function w.r.t. $\delta$ which operates on a set of words defined on the alphabet $\Sigma = \Sigma_Q \cup \Sigma_A$, and for all  $q \in \metq$ and $x \in \Sigma^*$:
\begin{align*}
    \delta^*: & \metq \times \Sigma^* \rightarrow \metq \\
    & \delta^*(q, ()) = q \\
    & \delta^*(q, xa) = \delta(\delta^*(q, x), a) \quad \text{if }  a \in \Sigma_A \\
    & \delta^*(q, xa) = \delta^*(q, x) \quad \text{if } a \in \Sigma_Q 
\end{align*}

A language $\metlang$ recognized by a DDFMA is defined as:
$$\metlang = \{w \in \Sigma^* \mid \delta^*(q_0, w) \in F\}$$

\end{definition}

Let us notice that the constraints $(1)$ and $(2)$ for the function $\bel$ translate that (1) if a transition does not exist in the meta-automaton, then the practitioner belief for the patient to 
be in the class associated with this transition cannot be different from 0, and (2) the sum of all beliefs associated with other transitions is equal to 1.  

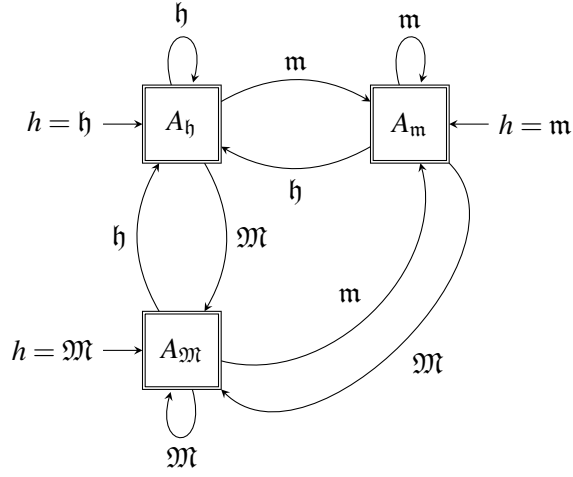
\begin{figure}
    \centering
    \begin{tikzpicture}[auto, node distance=3cm, on grid, >=stealth, state/.style={rectangle, draw, minimum size=1cm}]
    \node[state, initial, accepting, initial text={$h = \healthyc$}] (q0) {$A\Ah$};
    \node[state, right=of q0, initial,accepting, initial text={$h = \mediumc$}, initial where=right] (q1) {$A\Am$};
    \node[state, below=of q0, initial,accepting, initial text={$h = \strongc$}, initial where=left] (q2) {$A\As$};

    \path[->] (q0) edge[loop above] node {$\healthyc$} (q0);
    \path[->] (q0) edge[bend left] node {$\mediumc$} (q1);
    \path[->] (q1) edge[bend left] node {$\healthyc$} (q0);
    \path[->] (q1) edge[loop above] node {$\mediumc$} (q1);
    \path[->] (q2) edge[bend left] node {$\healthyc$} (q0);
    \path[->] (q0) edge[bend left] node {$\strongc$} (q2);
    \path[->] (q2) edge[bend right=60] node {$\mediumc$} (q1);
    \path[->] (q1) edge[bend left=90] node {$\strongc$} (q2);
    \path[->] (q2) edge[loop below] node {$\strongc$} (q2);
    \end{tikzpicture}
    \caption{Automaton of the experimental protocol \cite{healthinf25}}
    \label{fig:pnfa-protocol}
\end{figure}

For our medical application, we consider the three classes $\C = \{\healthyc,\mediumc,\strongc\}$ and the  DDFMA $\A = (\metq$, $\Sigma_{\C} $, $ \Sigma_{A},$ $ \delta_{exp},$ $ \prob_{exp},$ $ h,$ $ F)$, where
\begin{itemize}
    \item $\metq = \{A\Ah,A\Am,A\As\}$ is the set of states, which is composed by the three automata;
    \item $\Sigma_{\C} = \Sigma_{\Ah} \cup \Sigma_{\Am} \cup \Sigma_{\As}$ is the set of all symbols recognized by the automaton;
    \item $\Sigma_{A} = \C$: a symbol corresponds to a class that will be tested in the next state of the automaton 
    \item $\delta$ is given in Figure~\ref{fig:pnfa-protocol};
    \item for all $A_q \in \metq,$ for a word $w\in L_q$ recognized by $A_q$, for all $O \in \Sigma_{exp}$, the belief function $\bel_{exp}$ is such that  for all $q \in \metq, w \in L_{q}$, $x\in\Sigma_{A}$:
\begin{align*}
    & \bel_{q}(w,\healthyc,A\Ah)  =  \\
    & \left\{
    \begin{array}{ll}
         1 & \mbox{if } \Delta(w) < \Delta_{q}^{\healthyc} \\ 
       v_{q}^{\healthyc}+\frac{a_{q}^{\healthyc}}{b_{q}^{\healthyc}+\ c_{q}^{\healthyc} e^{\left(d_{q}^{\healthyc} \Delta(w) +z_{q}^{\healthyc}\right)}} & \mbox{otherwise}
    \end{array}
\right. \\
    & ~ & \\
    & \bel_{q}(w,\strongc,A\As)  =  \\
    & \left\{
    \begin{array}{ll}
         0 & \mbox{if } \Delta(w) < \Delta_{q}^{\strongc} \\ 
       v_{q}^{\strongc}+\frac{a_{q}^{\strongc}}{b_{q}^{\strongc}+\ c_{q}^{\strongc} e^{\left(d_{q}^{\strongc} \Delta(w) +z_{q}^{\strongc}\right)}} & \mbox{otherwise}
    \end{array}
    \right. \\
    & ~ & \\
    & \bel_{q}(w,\mediumc,A\Am)  =  1 - \bel_{q}(w,\healthyc,A\Ah) - \bel_{q}(w,\strongc,A\As)\\
\end{align*}

 \begin{table}
     \centering
     \begin{tabular}{|l||c|c|c|}
\hline
$q $ & $A\Ah$ & $A\Am$ & $A\As$ \\\hline
$\Delta_{q}^{\healthyc}$ & 2.016 & 0 & 0 \\\hline
$a_{q}^{\healthyc}$ & 0.5 & 1 & -0.1 \\\hline
$b_{q}^{\healthyc}$ & -3.6 & 1.2 & 0.1 \\\hline
$c_{q}^{\healthyc}$ & 1 & 0.3 & 0.1 \\\hline
$d_{q}^{\healthyc}$ & 0.7 & 2.2 & -2.4 \\\hline
$v_{q}^{\healthyc}$ & 0 & 0 & 1 \\\hline
$z_{q}^{\healthyc}$ & 0 & -1 & 1.1 \\\hline
$\Delta_{q}^{\strongc}$ & 6.256 & 0 & 3.769 \\\hline
$a_{q}^{\strongc}$ & 2.4 & -1 & -6.6 \\\hline
$b_{q}^{\strongc}$ & 2.1 & 1 & 0.4 \\\hline
$c_{q}^{\strongc}$ & -1 & 1.4 & 0.01 \\\hline
$d_{q}^{\strongc}$ & 0.24 & 0.8 & 1.6 \\\hline
$v_{q}^{\strongc}$ & 1 & 1 & 1 \\\hline
$z_{q}^{\strongc}$ & 0 & -6.3 & 0.4 \\\hline
\end{tabular}
     \caption{Factors of $\bel_q$, for all $q \in \metq$ \cite{healthinf25}}
     \label{tab:factors}
 \end{table}

    \item $h \in Q$ is the initial state, that corresponds to the initial hypothesis on the class in which the patient belongs to.

    \item $F = Q$: we consider each state to be an accepting state. 

\end{itemize}

 We defined a function $\Delta: L_q \rightarrow \mathbb{R}$ representing a \emph{confidence score} for the patient  not to belong to the class tested by the automata $\A_q$. The domain of this function is defined on $[0,m]$, with $m \in \mathbb{R}$ which is the max score (here $m = 10$). $0$ indicates that the patient did not do any mistake, while $m$ indicates  that the patient did $100\%$  mistakes or left the game, i.e., she did the $\theta$ action. 
We consider a factor for each action: $k_{\beta} = 1$ and $k_{\alpha} = 1$ since we aim to count the number of mistakes. However, since we have also $\theta$ and $\gamma$ as possible, to count these actions we associate a factor to each one. We consider a factor $k_{\gamma} = 0.2$ indicating that five $\gamma$ are equivalent to do one mistake. Finally, $k_{\theta} = 1\times 10^{9}$: if the patient leaves  the game, we consider it as $100\%$  mistakes.
Thus, $\Delta$ computes the number of mistakes (i.e., $\beta$ actions) and weights the number of waiting actions (i.e., $\gamma$ actions). To do so, we introduce a function for each $x \in \Sigma_{\C}$, $|.|_{x}:\Sigma_{\C}^{*} \rightarrow \mathbb{N} $ that counts the number of $x$ in a word $w \in \Sigma_{\C}^{*}$ and we note $|w|_{x}$ this number. The formal definition is the following one, for all words $w \in \Sigma_{\C}^{\star}$: 
$$\Delta(w)= \left\{
    \begin{array}{ll}
         m & \mbox{if } \theta \in w\\
        \frac{m \times(k_{\beta}\times|w|_{\beta}+k_{\gamma} \times |w|_{\gamma}+k_{\theta}\times|w|_{\theta})}{k_{\alpha}\times|w|_{\alpha}+k_{\beta}\times |w|_{\beta}+k_{\gamma} \times |w|_{\gamma}+k_{\theta}\times|w|_{\theta}} & \mbox{otherwise}
    \end{array}
\right.$$

\begin{figure}[h]
    \centering
    \includegraphics[width=1\textwidth,height=.4\textwidth]{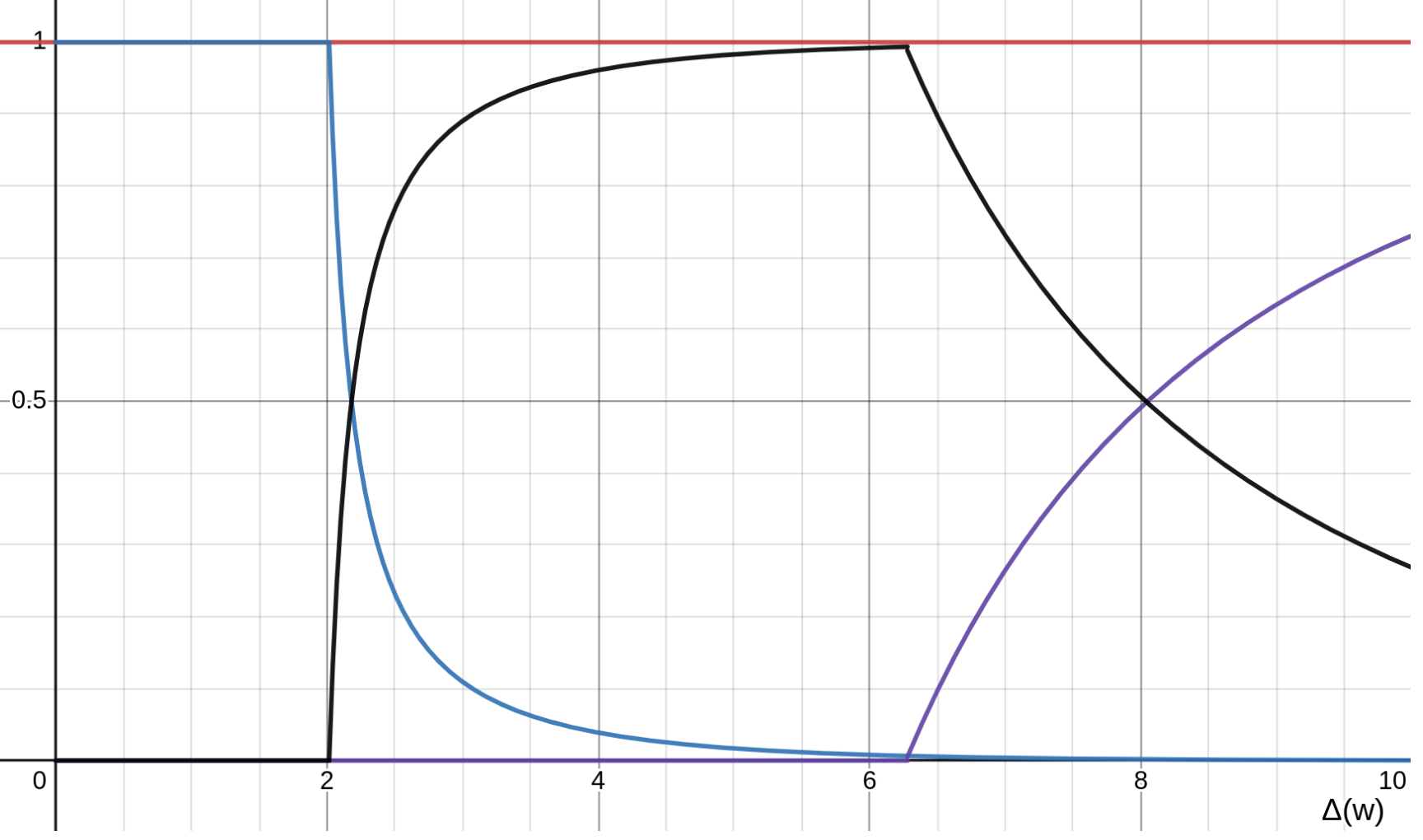}
    \caption{Evolution of $\bel_{A\Ah}$ in function of $\Delta(w)$ \cite{healthinf25}}
    \label{fig:mhealthy}
\end{figure}

\begin{figure}[h]
    \centering
    \includegraphics[width=1\textwidth,height=.4\textwidth]{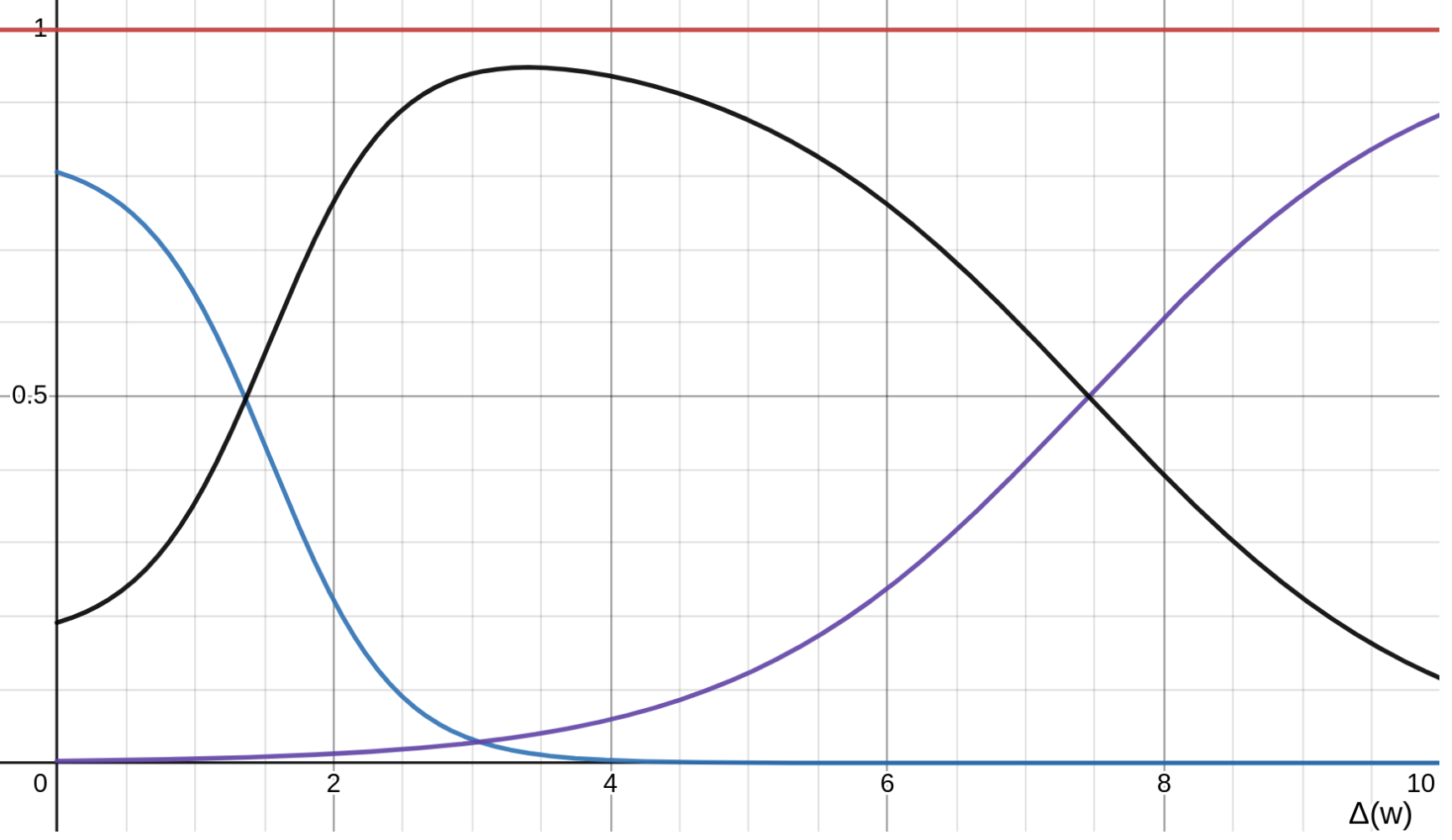}
    \caption{Evolution of $\bel_{A\Am}$ in function of $\Delta(w)$ \cite{healthinf25}}
    \label{fig:mmedium}
\end{figure}

\begin{figure}[h]
    \centering
    \includegraphics[width=1\textwidth,height=.4\textwidth]{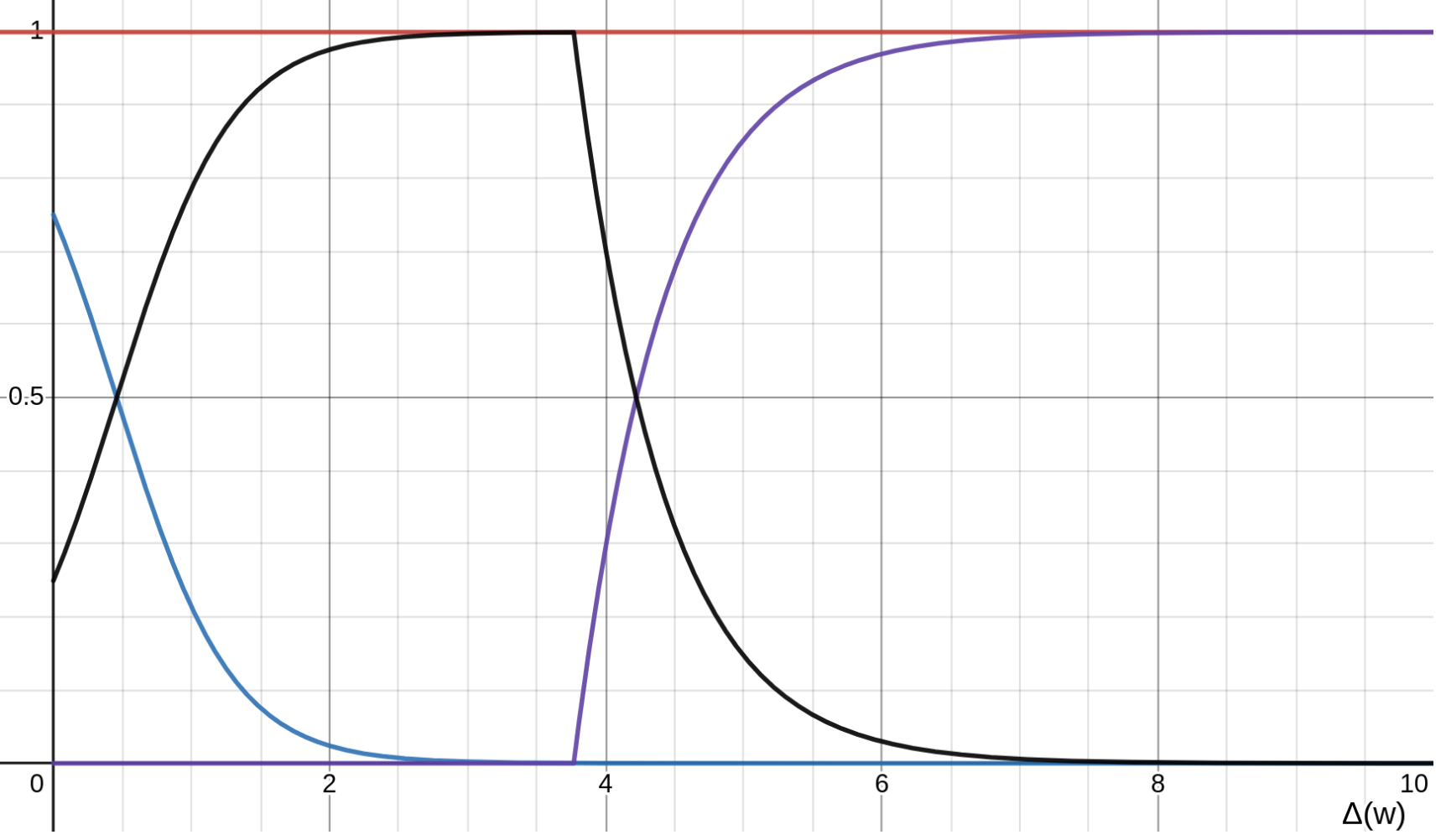}
    \caption{Evolution of $\bel_{A\As}$ in function of $\Delta(w)$ \cite{healthinf25}}
    \label{fig:mstrong}
\end{figure}

In Table~\ref{tab:factors}, we give the factors defined for each $\bel_q$. 
These factors considered in each $\bel_q$ have been computed based on  Table~\ref{tab:avg-prob} so that the output class from a test  corresponds to the class given by this table. Figure~\ref{fig:mhealthy},~\ref{fig:mmedium} and~\ref{fig:mstrong} represent the different belief functions for each automaton $A\Ah$, $A\Am$ and $A\As$, respectively. Thus,  to be considered in $\healthyc$, the patient has to produce at least $80\%$ of good answers (the value given in Table~\ref{tab:avg-prob} is 0.84). So, the delta is computed as $1-0.84 = 0.16$. This value corresponds to the abscissa of the first intersection between the green and the black curves in Figure~\ref{fig:mmedium}. The green curve represents the variation of the belief $\bel_{A\Am}(w,\healthyc,A\Ah)$, i.e., when we believe the patient could be in $\healthyc$. The black curve represents the variation of the belief $\bel_{A\Am}(w,\mediumc,A\Am)$, i.e., when we believe  the patient could be in $\mediumc$. This intersection between the green curve and the black one is approximately for a $\Delta(w) = 0.16$, it represents the moment where the practitioner considers the patient in $\mediumc$ and should do the test again. After $75\%$ mistakes, we believe that the patient could be in $\strongc$ and this is represented by the intersection between the black curve and the purple curve. The purple curve represents the variation  $\bel_{A\Am}(w,\strongc,A\As)$.  Figure~\ref{fig:mhealthy} represents the change of belief in function of the result $w$ of the patient when the test $A\Ah$ has been done. The green curve represents $\bel_{A\Ah}(w,\healthyc,A\Ah)$, the black curve $\bel_{A\Ah}(w,\mediumc,A\Am)$, and the purple curve $\bel_{A\Ah}(w,\strongc,A\As)$. After more than $20\%$  mistakes, the belief to be in $\healthyc$ decreases a lot to reach the intersection with the belief to be in $\mediumc$. Let us notice that, contrary to the green curve in Figure~\ref{fig:mmedium}, the belief to be in $\healthyc$ is slightly shifted to the right, since we consider the  test $A\Ah$ to be harder, and therefore tolerate the patient to make a few more errors. After $80\%$  mistakes, we start to believe the patient could belong to the class $\strongc$ and we would have to do the $\strongc$ transition. Figure~\ref{fig:mstrong} represents the change of belief in function of the result $w$ of the patient when the test $A\As$ has been done. 
We do not tolerate more than $13\%$  mistakes to be in class $\healthyc$. This is depicted by the intersection point between the green curve and the black curve. Let us notice that, since we applied the easiest test $A\As$, we cannot fully believe  the patient belongs to the class $\healthyc$ even if the patient does no mistake. This is represented by the fact that, for $\Delta(w) = 0$, $\bel_{A\As}(w,\healthyc,A\Ah) = 0.8$.

Figure~\ref{fig:pnfa-protocol} represents the experimental protocol as a DDFMA. The transitions represent the next test to apply. To know the next transition, i.e., the next test to apply, we compute a score from the current test with the $\Delta$ function. Then, this score is considered as an input for the belief function $\bel$. This function represents the beliefs of a clinician or a set of clinicians about the results obtained from the test. 

The first test to apply, i.e., the initial state, depends on the initial hypothesis $h$ we consider. Then, we start by the corresponding test $q$ to validate or reject $h$ and get a score; then this score is considered as an input for the function $\bel_q$ that returns the next test to apply. 

Let us admit the initial hypothesis is $h = \healthyc$, i.e., the patient is assumed to be in the class $\healthyc$. We then apply the test $A\Ah$: after the test is over, we have evaluated if the patient is in $\healthyc$ or if we have to reject the hypothesis. To compute the score we consider the $\Delta$ function previously defined,  which computes the number of actions that are not the right action (i.e., $\alpha$). 

If we could not reject the hypothesis, the evaluation is  $\healthyc$, which means we confirm the hypothesis as acceptable w.r.t. the test. If the result is $\mediumc$, it is highly possible that the patient is in $\mediumc$. The change of class will be suggested to practitioners.


\section{Formal validation} ~\label{sec:disc}
In this section, we apply the DDFMA framework to the serious game introduced in Section \ref{sec:case-study}, employing a PDFA-based representation to describe the expected behavior of each class of patients while playing the game. Firstly, we provide examples of PCTL properties to test the aptitude of the  models to display some interesting behaviours. 
Secondly, we apply LTL to the execution traces of the meta-automaton in order to define stopping conditions for the protocol. Thirdly, we give concrete decision-making capabilities of the meta-automaton and show 
some properties holding in our medical application, including examples of acceptable and unacceptable traces.

\subsection{Using PCTL to assess model probabilistic behavior}

The behaviour of the game for each class of patients is modelled with a PDFA that has the Markov Chain Property.  In order to use model-checking for our application, we define \emph{PDFA models} :

\begin{definition} 
Let $Atm$ be a set of symbols called \emph{atomic propositions}. We call $M = (A, \V)$ a \emph{PDFA model} iff $A$ is PDFA and  $\V: Atm \rightarrow 2^{Q}$ is a valuation function where $Q$ is the set of states of $A$. 
 We call $\M = (\A,\V)$ a \emph{DDFMA model} iff $\A$ is a DDFMA and $\V: Atm \rightarrow 2^{\metq}$ is a valuation function where $\metq$ is the set of states of $\A$.

\end{definition}

We define the following PDFA models: $M_{\healthyc} = (A\Ah, \V\Ah)$, $M_{\mediumc} = (A\Am, \V\Am)$, and $M_{\strongc} = (A\As, \V\As)$, for each automaton $A\Ah$,$A\Am$, and $A\As$. 
We consider  each action to be represented as an atomic proposition, denoted by $Atm = \{$ $\alpha$, $\beta$, $\gamma$, $\theta \}$, verified in the accessible state with the corresponding transition. 
Given a PDFA model $M \in \{M_{\healthyc},M_{\mediumc}, M_{\strongc}\}$,  if we access a state $s$ with the action $\beta$, then the atom $\beta$ is verified in $s$, denoted by $M, s \models \beta$. Each final state $f_1$ (resp. $f_2$) has a corresponding atom $a_1$ (resp. $a_2$), i.e., $ M, f_1 \models a_1$ and $M, f_2 \models a_2$. Given a model $M_c$, with $c \in \{\healthyc,\mediumc,\strongc\}$, and a state $s$ in $M_c$, we check if the state $s$ in $M_c$ satisfies the following PCTL useful property patterns.%
The properties given in this section have emerged after discussions with the clinicians at Claude Pompidou Institute.

\paragraph{Reachability of a final state.}

Is the probability of reaching a final state equal to 1? The PCTL formula to test is \texttt{P =1 [F (a1 or a2)]} 
which corresponds to the formal property : $M_c, s \models P_{=1} F (a_1 \vee a_2)$. 

\paragraph{Reachability of a precise configuration.}
Which is the probability of reaching a given state $s$? The PCTL formula to test is \texttt{P =? [F s]} 
which would correspond to find a $v \in \mathbb{R}$,  such that $M_c, s \models P_{=v} (F s)$. This property pattern is very useful to test  properties such as: Which is the probability for a patient to leave the
game before the maximum game duration? Which is the probability for a patient to start the
game and to interact with it $x$ times (not necessarily
consecutively) and not to interact with it $y$ times (not
necessarily consecutive)?

\paragraph{Reachability of a state  without violating some constraints.}
Is the probability to reach a certain state $d$ without violating some constraints $c$  greater than $0$? The PCTL formula to test is \texttt{P >0[c U d]} 
which corresponds to the formal property : $M_c, s \models P_{>0} (c U d)$.
An interesting example of property of this kind is: Does it happen that a patient  interacts
with the game until the end of the game without any
interruption?

\paragraph{Reachability of a state  without passing from another.}
Which is the probability to reach a state $d$ without passing from a state  $b$? The PCTL formula to test is  \texttt{P =?[(not b) U d]} which would correspond to find a $v \in \mathbb{R}$,  such that $M_c, s \models P_{=v} (\neg b U d)$.  
Useful examples of properties fitting this pattern is: Which is the probability for a patient to directly
choose the right picture, without choosing a wrong picture
before? Which is the probability for a patient to never
interact with the game until the end of the duration of the
game?

\paragraph{Concatenation of two actions.}
Is it true that every action of the kind $c$ is immediately followed by an action of the kind $d$? The PCTL formula to test is \texttt{P >0 [G(a-> Xb)]} 
which corresponds to the formal property : $M_c, s \models P_{>0} G[a \rightarrow Xb]$. An interesting example of property fitting this pattern is: Is any bad action of a patient followed by an inactivity?

\subsection{LTL analysis of patient experiences: protocol stop conditions}

In this section we consider the DDFMA $\A$ defined in Section~\ref{subsec:ddfma} and we note its corresponding DDFMA model $\M = (\A, \V)$. The valuation function $\V$ is such that each transition $t \in \C$  (where $\C =\{\healthyc, \mediumc, \strongc\})$ is associated with an atomic proposition verified in a state $q \in \metq$. For instance, if we do a transition $\mediumc$ to reach a state $q_m$, then the corresponding atomic proposition $\mediumc$ is verified in $q_m$, i.e., $\M, q_m \models \mediumc$.

\subsubsection{Oscillating behaviour}

If we detect a trace $\tau = (q_0, q_1, \ldots,q_n) \in \metq^{n+1}$  showing that the patient alternated between different classes, then we want to stop the protocol.
Hereafter we provide some examples of traces we aim to detect by providing a regular expression: $(\mediumc\text{ .}\ast \strongc)^{4}$ (resp. $(\healthyc\text{ .}\ast\strongc)^{4}$, $(\healthyc\text{ .}\ast\mediumc)^{4}$),  i.e., the patient oscillates between $\mediumc$  and $\strongc$ (resp. $\healthyc$ and $\strongc$, $\healthyc$ and $\mediumc$) four consecutive times. Here "." denotes any character of $\C$ and $\ast$ is the quantification "zero or more occurrences". 

Let $q_0$ be the first state of a trace $\tau$. $q_0$ verifies the stop condition if it satisfies the following LTL formula: \texttt{$\mediumc$ and F ($\strongc$ and F ($\mediumc$ and F ($\strongc$ and  F($\mediumc$ and F ($\strongc$ and F ($\mediumc$ and F $\strongc$ ))))))}. 


\subsubsection{The permitted number of tests has been exceeded}

We consider that a patient will not do more than 10 tests. Given an initial state $q_0$ for a trace $\tau$, $q_0$ verifies this stop condition if it satisfies the LTL formula: \texttt{$X^{10}$ true and not $X^{11}$ false}, where  $X^{n}$ is a shortcut for $X...X$ with $n$ occurrences of $X$.  

\subsubsection{Reaching a steady-state condition}

If a patient stays in the same class for at least 3 tests, we stop the test.
For instance, given a trace $\tau = (q_0,\ldots,q_n)$, a state $q_{n-2}$ verifies this stop condition for the healthy class if it verifies the LTL formula: \texttt{$\healthyc$ and X($\healthyc$ and X $\healthyc$)}.%
\subsection{Acceptable and unacceptable traces}

Some traces are unacceptable. For instance, if the patient gives wrong answers but he is considered as $\healthyc$, then we do not want her to stay in the class $\healthyc$. In the same way,  if the patient has a very positive outcome but is considered as $\mediumc$, then we would like 
to check in a more accurate way if she could be classed as $\healthyc$. But we allow a mitigated outcome to make the patient stay in the same state. In the following, given a DDFMA model $\M = (\A, \V)$, we formally show that our belief functions are compatible with these properties.

\subsubsection{An extremely poor outcome in $A\Ah$ cannot classify a patient as $\healthyc$}

A trace in the DDFMA looks like $\tau = \healthyc \color{blue}\beta \gamma * \beta \gamma * \beta   \color{black} \healthyc $. This trace recognized by a DDFMA means that we initialize the DDFMA in the state $A\Ah$ to make the patient perform this test. In this test, the word recognized by the automaton $A\Ah$ 
is $\beta \gamma * \beta \gamma * \beta $. This trace signifies that the patient only does wrong answers $\beta$, and may wait between actions, i.e., $\gamma *$.

We formally show that, according to the DDFMA model defined in Section~\ref{subsec:ddfma}, this is impossible. 
Let consider a word $w = \beta \gamma \gamma \beta \beta \beta \gamma \beta^5 \gamma \beta $ recognized by  the test $A\Ah$, i.e., this is the execution of actions of the patient. Here obviously  $\Delta(w) = m = 10$. Thus according to Figure~\ref{fig:mhealthy}, the belief $\bel_{A\Ah}(w,\strongc,A\As) = 0.72$ and $\bel_{A\Ah}(w,\mediumc,A\Am) = 0.28$. Since the belief of belonging to $\strongc$ is stronger than belonging to $\mediumc$, the automaton should propose a transition towards $A\As$ and so $\tau$ cannot be verified in this configuration for our DDFMA.

\subsubsection{A very Good outcomes in $A\As$ can classify a patient as $\healthyc$}

Let consider that a patient does $100\%$ of $\alpha$ in the automaton $A\As$. A word recognized by $A\As$ could be $w = \alpha^{10}$. Thus, $\Delta(w) = 0$ and $\bel_{A\As}(w,\healthyc,A\Ah) = 0.75$ if we look at Figure~\ref{fig:mstrong}. In such situation an acceptable trace would be $\strongc \color{blue} w \color{black} \healthyc$.  


\subsubsection{A medium outcomes in $A\Am$ can classify a patient as $\mediumc$}
 A medium outcome in $A\Am$ corresponds to all words $w$ recognized by $A\Am$ such that $\Delta(w) \in [1.364,7.45]$. For instance a word $w = (\alpha \beta)^{5}$ have a $\Delta(w) = 5$ and so $\bel_{A\Am}(w,\mediumc,A\Am) = 0.8765$.  In such situation an acceptable trace by the DDFMA would be $\mediumc w \mediumc$.

\section{Conclusion and Future Work} ~\label{sec:conclu}
Artificial intelligence continues to transform the medical field, offering innovative solutions to enhance patient care and diagnostic accuracy \cite{lu2021application,panagoulias2022microservices}. In this work, we introduced a framework to model the behavior of Alzheimer’s patients during multiple sessions of serious games. Our approach serves two primary purposes: (i) to support medical diagnosis by analyzing patient performance during gameplay, and (ii) to assist practitioners in determining the appropriate next step, whether it be a new game or an adjusted difficulty level, based on the patient’s previous session.

A key strength of the proposed methodology is its adaptability, making it applicable to other medical protocols, such as diagnosing and training children with attention disorders. The framework was developed in close collaboration with clinicians at Claude Pompidou Institute (Nice, France), ensuring its alignment with practical needs and clinical workflows.

Before proceeding with implementation, we aim to present a comprehensive theoretical model to practitioners for feedback and refinement. Subsequently, we plan to develop the model using the Probabilistic Model Checker PRISM \cite{KNP11CAV,LD2021SCP,DGRR18Bioinfo}, providing clinicians with a user-friendly, automated tool. For simplicity, the current framework addresses a single game and difficulty level; however, it can be extended to include multiple games targeting distinct cognitive functions, such as memory or inhibitory control, with varying levels of complexity.

Moreover, the formal approach outlined in this work paves the way for automating medical protocols and providing dynamic feedback to practitioners regarding the progression of a patient’s cognitive state. The proposed meta-automaton continuously evaluates practitioners’ confidence levels in the diagnosis based on the consistency and evolution of patient performance across sessions. This adaptive mechanism not only refines diagnostic accuracy over time but also personalizes cognitive training, adjusting game complexity based on prior performance. Thus, the proposed framework positions serious games as a powerful dual-purpose tool for both diagnosis and personalized training in the context of neurodegenerative diseases.

\bibliography{references} 

\end{document}